\documentclass{emulateapj}
\pdfoutput=1
\usepackage{natbib}
\usepackage{times}
\usepackage{graphicx}
\usepackage{amsmath}

\usepackage[colorlinks=true,citecolor=black,filecolor=blue,linkcolor=blue,urlcolor=blue,pdftex]{hyperref}

\begin{document}

\journalinfo{UCB-NPAT-13-003, NT-LBNL-13-003}
\shorttitle{GRBs and the Faintest of Galaxies at High $z$}
\shortauthors{Kistler, Y{\"u}ksel, \& Hopkins}

\title{The Cosmic Star Formation Rate from the Faintest Galaxies in the Unobservable Universe}

\author{Matthew D.~Kistler\altaffilmark{1,2,4},
Hasan Y{\"u}ksel\altaffilmark{1}, and
Andrew M. Hopkins\altaffilmark{3}
}

\altaffiltext{1}{Lawrence Berkeley National Laboratory, Berkeley, CA 94720, USA}
\altaffiltext{2}{Department of Physics, University of California, Berkeley, CA 94720, USA}
\altaffiltext{3}{Australian Astronomical Observatory, P.O.\ Box 915, North Ryde, NSW 1670, Australia}
\altaffiltext{4}{Einstein Fellow}



\begin{abstract}
Observations of high-$z$ galaxies and gamma-ray bursts now allow for empirical studies during reionization.  However, even deep surveys see only the brightest galaxies at any epoch and must extrapolate to arbitrary lower limits to estimate the total rate of star formation.  We first argue that the galaxy populations seen in LBG surveys yield a GRB rate at $z \!>\! 8$ that is an order of magnitude lower than observed.  We find that integrating the inferred UV luminosity functions down to $M_{\rm UV} \!\approx\! -10$ brings LBG- and GRB-inferred SFR density values into agreement up to $z \!\sim\! 8$.  GRBs, however, favor a far larger amount of as yet unseen star formation at $z \!\gtrsim\! 9$.  We suggest that the SFR density may only slowly decline out to $z \!\sim\! 11$, in accord with {\it WMAP} and {\it Planck} reionization results, and that GRBs may be useful in measuring the scale of this multitude of dwarf galaxies.
\end{abstract}



\section{Introduction}
Searches for the earliest galaxies and gamma-ray bursts have advanced in step, with GRBs observed at $z\!\approx\!8.2$ \citep{Salvaterra09,Tanvir09} and $z\!\approx\!9.4$ \citep{Cucchiara:2011pj} and galaxy candidates seen at $z\!\sim\!11$ \citep{Coe:2012wj} and $z\!\sim\!12$ \citep{Ellis:2012bh}.  GRBs can (briefly) probe large volumes of the universe extending to high $z$, while Lyman Break Galaxy (LBG) surveys can make deep observations of narrow regions.  These allowed for initial assessments of the star formation rate density ($\dot{\rho}_*$) in the reionization era \citep{Kistler:2009mv,Bouwens:2009at,Yan:2009qa}.

\begin{figure}[b!]
\includegraphics[width=3.45in,clip=true]{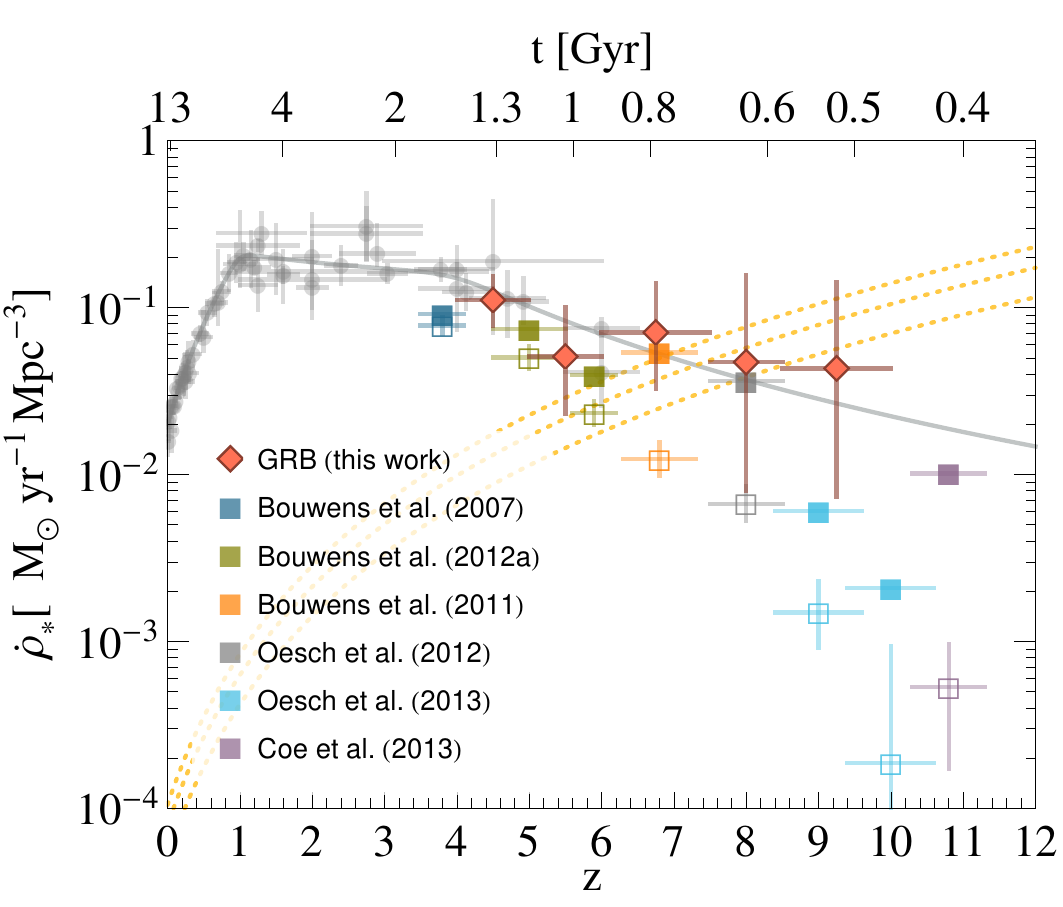}
\caption{The cosmic star formation history.  Low-$z$ data ({\it circles}) are from the compilation of \citet{Hopkins:2006bw}.  The {\it diamonds} are our values obtained using \textit{Swift} gamma-ray bursts.  The {\it open squares} show the result of integrating the LBG UV luminosity functions down to the lowest measured value, $M_{\rm vis}$, while the {\it solid  squares} use $M_{\rm cut}\!=\!-10$ (see Table~\ref{tab:params}).  All assume a Salpeter IMF.  For comparison, we show the critical $\dot{\rho}_*$ from \citet{Madau et al.(1999)} for $\mathcal{C}/f_{\rm esc}\!=\!40,\,30,\,20$ ({\it dotted lines}, top to bottom).
\label{SFH}}
\end{figure}

Typically, the measured galaxy luminosity function (LF) at a given $z$ is integrated to some arbitrary lower limit to estimate the total $\dot{\rho}_*(z)$.  At low $z$, this cutoff is not vital, since the faint-end slope $\alpha$ in the common $dn/dL\!\propto\!L^{\alpha}\,e^{-L/L_*}$ form of the LF is shallower than the divergent $\alpha\!=\!-2$, so that integration could even be taken down to $L\!=\!0$.  At high $z$, LBG observations have revealed that the faint end of the UV LF becomes quite steep (e.g.,  \citealt{Bouwens07,Bouwens:2010gp,Bouwens:2011xu,Reddy2009,Oesch:2011cj,Finkelstein:2012rk}), reaching $\alpha\!\lesssim\!-2$ at $z\!\gtrsim\!7$, so that the choice of cutoff becomes crucial at these epochs (see Fig.~\ref{SFH}).

Surveys now suggest that the $\dot{\rho}_*$ from bright galaxies declines strongly at high-$z$ \citep{Ellis:2012bh,Oesch:2013pt}.  The total $\dot{\rho}_*$ at $z\!\gtrsim\!7$ may well be dominated by the contribution of unseen faint galaxies, which could be influenced by novel physics (e.g., \citealt{Kuhlen:2011dt,Jaacks:2013hx}).  The establishment of gamma-ray bursts as an outcome of the core collapse of massive, and thus short-lived, stars \citep{Stanek:2003tw,Hjorth} implies that they trace distant star formation \citep{Totani,Wijers:1998,Lamb,Porciani,Yuksel:2008cu} and could be used to probe the total instantaneous star formation history.

In light of new data obtained through the efforts of {\it Swift} \citep{Gehrels:2004am} GRB searches and {\it HST} WFC~3 galaxy surveys, we re-examine $\dot{\rho}_*$ as indicated by gamma-ray bursts in connection to the high-$z$ UV LF.  GRBs at low $z$ are observed to occur predominantly in metal-poor \citep{Stanek:2006gc,Graham:2012ga}, sub-$L_*$ galaxies \citep{Fynbo:2003sx, Le Floc'h:2003yp, Fruchter}.  \citet{Kistler:2009mv} concluded that GRBs likely trace this faint population at high $z$, in good agreement with a lack of galaxies found in deep searches for high-$z$ GRB hosts \citep{Tanvir:2012nv,Trenti:2012gc}, and that such galaxies could have generated a sufficient $\dot{\rho}_*$ to account for cosmic reionization.

We first reassess the evolution of GRBs relative to the SFR history at lower $z$, making use of the increased number of detections made by {\it Swift} and GRB observers.  Using this calibration, we find it to be quite unlikely to have seen two $z\!>\!8$ GRBs, or even one at $z \!\gtrsim\! 9$, from the populations of galaxies directly observed in surveys.  We determine that GRB and LBG data together imply an increasing abundance of faint galaxies with $z$.  This suggests the number of diminutive galaxies during reionization declines with time, consistent with the rapid growth in mass expected from this early star formation epoch.  We further discuss evidence in favor of GRBs at $z\!>\!10$ existing in sufficient numbers for a possible imminent detection and how GRBs may be used in establishing a characteristic star formation rate scale in high $z$ galaxies.


\section{Swift Gamma-ray Bursts}

\begin{figure}[t!]
\includegraphics[width=3.425in,clip=true]{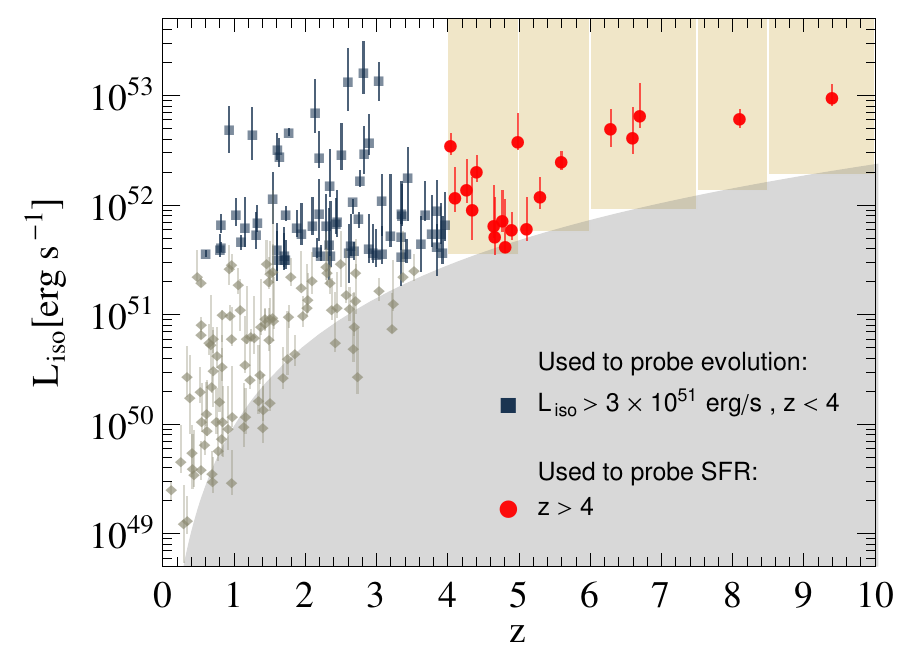}
\caption{The luminosity-redshift distribution of 184 \textit{Swift} GRBs that we obtain using the updated catalog of \citet{Butler:2007hw,Butler:2009nx}, with an effective detection threshold illustrated by the curved shaded region.  The dark blue squares are the 67 GRBs at $z\!<\!4$ that we use to determine the cosmic evolution of the GRB rate relative to the star formation rate.  The circles at $z\!>\!4$ are separated into subsamples used to estimate the high-$z$ SFR density.
\label{Lz}}
\end{figure}

To infer the cosmic SFR density ($\dot{\rho}_*$) from the GRB rate, we need to understand their connection quantitatively.  Following \citet{Kistler}, we calculate the expected distribution of GRBs in $z$ using their comoving rate, $\dot{n}_{\rm GRB}(z)\!=\!\mathcal{E}(z)\!\times\!\dot{\rho}_*(z)$, where $\mathcal{E}(z)$ is the redshift-dependent fraction of stars that produce GRBs.  We then account for the likelihood to obtain a redshift ($0\!<\!F(z)\!<\!1$), the fraction of GRBs that are observable due to beaming ($\left\langle f_{\rm beam} \right\rangle$), and $dV/dz$\footnote{$dV/dz\!=\!4\pi\,(c/H_0)\,d_c^2(z)/\sqrt{(1+z)^3\,\Omega_{\rm m}+\Omega_\Lambda}$, where $d_c$ is the comoving distance, $\Omega_{\rm m}\!=\!0.3$, $\Omega_\Lambda\!=\!0.7$, and $H_0\!=\!70$~km/s/Mpc.} through
\begin{equation}
  \frac{d\dot{N}}{dz} = F(z)	\frac{\mathcal{E}(z)\, \dot{\rho}_*(z)}{\left\langle f_{\rm beam}\right\rangle} \frac{dV/dz}{1+z}\,.
\label{dndz}
\end{equation}
By considering GRBs with sufficient rest-frame luminosity to be visible throughout a given range of redshifts, keeping $F(z)$ roughly constant, we avoid a detailed comparison between the GRB LF and detection thresholds.  We use the parametrization $\mathcal{E}(z)\!=\!\mathcal{E}_0 (1+z)^{\eta}$, with $\mathcal{E}_0$ a constant that converts $\dot{\rho}_*(z)$ to a GRB rate in the same luminosity range.

Based on 36 bright {\it Swift} GRBs with redshifts, \citet{Kistler} found that a direct correlation with the SFR history ($\eta\!=\!0$) was excluded at the $\sim\,$95\% level.  This increased to $\sim\,$99\% in \citet{Kistler:2009mv}.  In Fig.~\ref{Lz}, we show the average rest-frame GRB luminosities, $L_{\rm iso}\!=\!E_{\rm iso}/[T_{90}/(1+z)]$ found using the isotropic equivalent $1\!-\!10^4$~keV energy release ($E_{\rm iso}$) for 184 GRBs with $T_{90}\!>\!2$~s from \citet{Butler:2007hw,Butler:2009nx}\footnote{Up to Jan.~1, 2013; see http://butler.lab.asu.edu/Swift.}.

As in \citet{Kistler}, we consider GRBs in $0\!<\!z\!<\! 4$ for comparison with SFR measurements.  While \citet{Kistler} used a cutoff of $L_{\rm iso}\!>\!10^{51}$~erg~s$^{-1}$, which was also used in a number of subsequent studies (e.g., \citealt{Kistler:2009mv,Robertson:2011yu}), this cut likely under-counts GRBs at $z\!\gtrsim\!3$, as seen in Fig.~\ref{Lz}.  This would result in systematically under-representing any evolution present in $\mathcal{E}(z)$ (e.g., \citealt{Robertson:2011yu} found $\eta \!=\! 0.5$).

The greater number of events now available allows for an improved assessment.  We rather use $L_{\rm iso}\!>\!3\!\times\!10^{51}$~erg~s$^{-1}$, based on the approximate detection threshold at $z\!\approx\!4$ in the GRB data in Fig.~\ref{Lz}.  Using a Monte Carlo method to sample $10^4$ realizations of redshift distributions, we find that the piecewise \citet{Hopkins:2006bw} $\dot{\rho}_*(z)$ fit alone is now incompatible with the GRB data at $\gtrsim\!99\%$.  This analysis presently suggests $\eta\!\approx\!1.2$.  This evolution seems most naturally explained due to cosmic metallicity, although it must be taken into account in relating GRBs to star formation regardless of the origin.  As a conservative measure, we assume that this evolution persists in this form to higher redshifts.


\section{How many high-z GRBs should we have expected?}

\citet{Yuksel:2008cu} presented a technique for estimating $\dot{\rho}_*$ via high-$z$ GRBs by using GRB and SFR data spanning $1\!<\!z\!<\!4$ as calibration for comparing bursts with luminosities above a given threshold value.  This method makes use of ratios to handle common parameters that would be difficult to determine on their own.  Using the piecewise $\dot{\rho}_*(z)$ of \citet{Hopkins:2006bw}, we start by finding the ``expected'' number of GRBs in $1\!<\!z\!<\!4$, where $\dot{\rho}_*(z)$ is fairly flat, as
\begin{eqnarray}
\mathcal{N}_{1-4}^{\rm exp}
  & = & \Delta t \frac{\Delta \Omega}{4\pi} \int_{1}^{4} dz\,  F(z) \, \mathcal{E}(z)  \frac{\dot{\rho}_*(z)} {\left\langle f_{\rm beam}\right\rangle} \frac{dV/dz}{1+z} \nonumber \\
  & = & \mathcal{A} \, \int_{1}^{4} dz\, \dot{\rho}_*(z)\, (1+z)^{\eta} \, \frac{dV/dz}{1+z}\,,
\label{N1-4}
\end{eqnarray}
in which $\mathcal{A}\!=\!{\Delta \Omega\,\Delta t\,\mathcal{E}_0\,F_0}/4\pi{\left\langle f_{\rm beam}\right\rangle}$ is based on sky coverage ($\Delta\Omega$), observing time ($\Delta t$), and a GRB luminosity cut.

Using the average $\dot{\rho}_*(z)$ in the range $z_1\!-\!z_2$, $\left\langle\dot{\rho}_*\right\rangle_{z_1-z_2}$, and taking the measured counts, $\mathcal{N}_{1-4}^{\rm obs}$, in lieu of $\mathcal{N}_{1-4}^{\rm exp}$, we obtain
\begin{eqnarray}
\mathcal{N}_{z_1-z_2}^{\rm exp}
  & = &  \left\langle \dot{\rho}_* \right\rangle_{z_1-z_2} \mathcal{A} \, \int_{z_1}^{z_2} dz\, (1+z)^{\eta} \, \frac{dV/dz}{1+z} \nonumber \\
  & = &  \left\langle \dot{\rho}_* \right\rangle_{z_1-z_2} \mathcal{N}_{1-4}^{\rm obs} \frac{\int_{z_1}^{z_2} dz\, \frac{dV/dz}{1+z} (1+z)^\eta}{\int_{1}^{4} dz\, \frac{dV/dz}{1+z} \dot{\rho}_*(z)\, (1+z)^\eta} .
\label{Nz1-z2}
\end{eqnarray}

\begin{figure}[t!]
\centering\includegraphics[width=3.4in,clip=true]{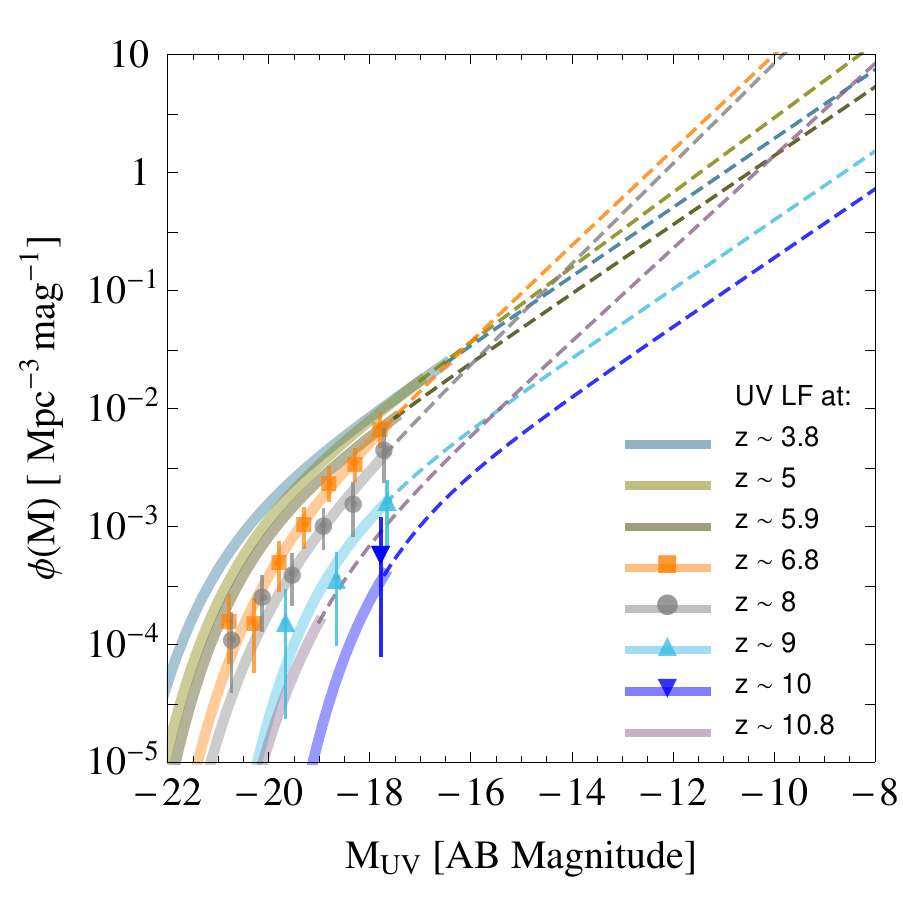}
\caption{The UV luminosity function at high redshift.  Shown are Schechter function fits for a range of redshifts (with values and references given in Table~\ref{tab:params}), along with measurements at $z\!\sim\!7$ \citep{Bouwens:2010gp}, $z\!\sim\!8$ \citep{Oesch:2011cj}, and $z\!\sim\!9$, 10 \citep{Oesch:2013pt}.  The solid bands extend down to the values used as integration cutoffs to estimate $\dot{\rho}_*$ in visible galaxies, $M_{\rm vis}$, with dashed lines extrapolating down past $M\!=\!-10$.\\
\label{LF}}
\end{figure}

We use Eq.~(\ref{Nz1-z2}) to first determine how many GRBs would be expected to arise from the populations of galaxies directly observed at $z\!\gtrsim\!7$ (as in Fig.~\ref{LF}).  In Fig.~\ref{SFH}, we show the result of integrating the quoted Schechter function UV LF of each LBG set down to the lowest measured data point for each (rather than the commonly used $M_{\rm UV}\!=\!-17.7\!=\!0.05\,L^*_{z=3}$), given as $M_{\rm vis}$ in Table~\ref{tab:params}.  The resulting $\dot{\rho}_*$ values ({\it open squares}) assume their dust corrections, a Salpeter IMF, and the original relative error bars.  We then bin GRBs in redshift ranges that roughly match those of the LBG data for a straightforward comparison.  Since GRBs within $1\!<\!z\!<\!4$ are used for calibration, it is important to account for the GRB luminosity threshold, which depends on $z$.  We use the values shown in Fig.~\ref{Lz} at the center of each of the $z$ bins as the lower cuts.

The expected GRB counts obtained from these $\dot{\rho}_*$ values are $\sim\!0.6$ for $z\!=\!6\!-\!7.5$, $\sim\!0.1$ for $z\!=\!7.5\!-\!8.5$, and $\sim\!0.03$ for $z\!=\!8.5\!-\!10$, significantly below the 3, 1, and 1 GRBs observed in these ranges, respectively.  The calculation already includes a factor of $\sim\,$15 enhancement in the GRB rate at $z\!=\!9$ as compared to $z\!=\!0$.  To bring these counts into agreement by adjusting this factor would require GRB production to become very efficient, very rapidly, with $\sim\,$50\% of all massive stars needing to end as GRBs at $z\!\sim\!9$ (assuming a 1/1000 local fraction).  The observation of a host brighter than $M_{\rm vis}$ at $z\!\gtrsim\!6$ would not be surprising (although this assumes all galaxies to be equally likely to yield a GRB, while those observed are the most metal enriched).  However, most of these GRBs should thus arise from a population of galaxies not yet represented in the LBG data sets, in accord with host luminosity limits \citep{Tanvir:2012nv,Trenti:2012gc}.

%
\begin{deluxetable*}{lccccccccl}[t!]
\tabletypesize{\scriptsize}
\tablecaption{\label{tab:params}}
\tablewidth{17cm}
\startdata
 \multicolumn{9}{c}{Star formation rate densities from integrating each luminosity function down to the faintest measured magnitude $M_{\rm vis}$ or $M_{\rm cut}\!=\!-10$.} \\
 \multicolumn{9}{c}{Values in parentheses are fixed from lower-$z$ data, while brackets indicate an empirical relation.} \\
\hline\hline \vspace{-0.2cm}\\
$z$	& Source 				& $\phi_*$ 			& $M_*$					& $\alpha$			& $M_{\rm vis}$& $\dot{\rho}_{*, {\rm vis}}$& $M_{\rm cut}$& $\dot{\rho}_{*,{\rm cut}}$ \\
	&					& ($\times 10^{-3}\,$Mpc$^{-3}$)&					&					&			& ($\times 10^{-2}\,$)\tablenotemark{a}& & ($\times 10^{-2}$)\tablenotemark{a} \\ \hline
3.8	& \citet{Bouwens07}			& $1.3 \pm 0.2$					& $-20.98 \pm 0.10$			& $-1.73 \pm 0.05$				& $-16.5$	& $7.8^{+0.9}_{-0.8}$				& $-10$		& $9.2$ \\
5.0	& \citet{Bouwens:2011xu}& $1.4^{+0.7}_{-0.5}$		& $-20.60 \pm 0.23$			& $-1.79 \pm 0.12$				& $-17$		& $5.0^{+0.8}_{-0.7}$				& $-10$		& $7.4$ \\
6.0	& \citet{Bouwens:2011xu}& $1.4^{+0.6}_{-0.4}$		& $-20.24 \pm 0.19$			& $-1.74 \pm 0.16$				& $-17.5$	& $2.3^{+0.4}_{-0.3}$				& $-10$		& $3.9$ \\
6.8	& \citet{Bouwens:2010gp}& $0.86^{+0.70}_{-0.39}$& $-20.14 \pm 0.26$			& $-2.01 \pm 0.21$				& $-17.7$	& $1.2^{+0.3}_{-0.3}$				& $-10$		& $5.4$ \\
8.0	& \citet{Oesch:2011cj}	& $0.50^{+0.70}_{-0.33}$& $-20.04^{+0.44}_{-0.48}$& $-2.06^{+0.35}_{-0.28}$& $-17.7$& $0.67^{+0.18}_{-0.14}$ 		& $-10$		& $3.6$ \\
9.0	& \citet{Oesch:2013pt}	& ($1.1$)								& $-18.8 \pm 0.3$				& $(-1.73)$								& $-17.7$	& $0.15^{+0.08}_{-0.06}$		& $-10$		& $0.60$ \\
10.0& \citet{Oesch:2013pt}	& ($1.1$)								& $-17.7 \pm 0.7$				& $(-1.73)$								& $-17.7$	& $0.019^{+0.075}_{-0.016}$ & $-10$		& $0.21$ \\
10.8& \citet{Coe:2012wj}		& ($0.43^{+0.35}_{-0.21}$)& [$-19.42$]& ($-1.98^{+0.23}_{-0.22})$					& $-19$		& $0.053^{+0.043}_{-0.036}$ & $-10$		& $1.0$ \\
\hline\vspace{-0.3cm}
\enddata
\tablenotetext{a}{in units of $M_\odot\,$yr$^{-1}\,$Mpc$^{-3}$}
\end{deluxetable*}
%


\section{The Star Formation History at the Highest Redshifts}

\citet{Kistler:2009mv} suggested that galaxies not currently detectable directly through their stellar emission likely dominate $\dot{\rho}_*$ at high $z$, with GRBs being the best near-term probes of their contribution.  We now use the latest GRB observations to better estimate this $\dot{\rho}_*$.  Using Eqs.~(\ref{N1-4}) \& (\ref{Nz1-z2}),
\begin{equation}
\left\langle \dot{\rho}_* \right\rangle_{z_1-z_2} = 
\frac{\mathcal{N}_{z_1-z_2}^{obs}}{\mathcal{N}_{1-4}^{obs}} \frac{\int_{1}^{4} dz\, \frac{dV/dz}{1+z} \dot{\rho}_*(z)\, (1+z)^\eta}{\int_{z_1}^{z_2} dz\,
\frac{dV/dz}{1+z} (1+z)^\eta}\,,
\label{zratio}
\end{equation}
where we use the same redshift ranges and luminosity cuts as above, with the resulting GRB subsamples displayed in Fig.~\ref{Lz}.

Fig.~\ref{SFH} displays our resulting $\dot{\rho}_*$ values.  For $z\!=\!4\!-\!5$, $5\!-\!6$, $6\!-\!7.5$, $7.5\!-\!8.5$, and $8.5\!-\!10$, $\dot{\rho}_*\!=\!0.111^{+0.044}_{-0.033}$, $0.051^{+0.049}_{-0.028}$, $0.071^{+0.069}_{-0.039}$, $0.047^{+0.11}_{-0.039}$, and $0.043^{+0.099}_{-0.036}$ $M_\odot\,$yr$^{-1}\,$Mpc$^{-3}$, respectively (with error bars using 68\% Poisson confidence intervals for the number of events in each bin).  If our assumption of continuing evolution is overly aggressive, perhaps due to a low-metallicity saturation, then the GRB-based $\dot{\rho}_*$ values could be increased by a factor reaching $\sim\,$2.  These estimates also assume a uniform detection efficiency in $z$.  Additional possible effects, due to choice of $L_{\rm iso}$ cutoff, luminosity estimator, or variation in $\eta$, were discussed in \citet{Yuksel:2008cu} and \citet{Kistler:2009mv}.  In Fig.~\ref{SFH}, we show our parametrization of $\dot{\rho}_*(z)$, using $\dot{\rho}_{0}\!=\!0.02\,M_\odot$~yr$^{-1}$~Mpc$^{-3}$,
\begin{eqnarray}
    \!\!\!\!\!  \dot{\rho}_*(z)
      & = &  \dot\rho_0 \left[(1 + z)^{{a}{\zeta} } + \left(\frac{1 + z}{B}\right)^{{b}{\zeta}} \!+ \left(\frac{1 + z}{C}\right)^{{c}{\zeta} } \, \right]^{1/\zeta} \!\!\!,
\label{sfhfit}
\end{eqnarray}
with slopes $a\!=\!3.4$, $b\!=\!-0.3$, and $c\!=\!-2.5$, breaks at $z_1\!=\!1$ and $z_2\!=\!4$ corresponding to $B\!=\!(1\!+\!z_1)^{1-a/b}\!\simeq\!5160$ and $C\!=\!(1\!+\!z_1)^{(b-a)/c}(1\!+\!z_2)^{1-b/c}\!\simeq\!11.5$, and $\zeta\!=\!-10$.

\section{Rise of the Dwarfs}

Comparing the GRB $\dot{\rho}_*$ values to those inferred from the visible galaxy population, we see a fair agreement at $z\!\sim\!4$, although a divergence develops with increasing $z$.  We attribute this to an increasing lack of sensitivity to the existence of faint galaxies.  We take an additional step of integrating the UV LFs down to a much fainter limiting magnitude.  For simplicity, we choose a $z$-independent cut of $M_{\rm cut}\!=\!-10$.  These are shown in Fig.~\ref{SFH} ({\it solid squares}) for the values given in Table~\ref{tab:params} (we have again used their dust corrections and not attempted to vary the LF parameters or reassess the $\dot{\rho}_*$ error bars).

Up to $z\!\sim\!8$, we see that this brings the LBG-based results in line with those from GRBs.  This fails at $z\!\sim\!9$, though, since even integrating the LF assumed by \citet{Oesch:2013pt} to $L\!=\!0$ falls well short.  Either we were lucky to see a $z\!\approx\!9.4$ GRB, the GRB was at a much lower $z$ than the photometric $z \!\approx\! 9.4$ of \citet{Cucchiara:2011pj}, or the $\dot{\rho}_*$ from the UV LF is too low for this regime.  We concentrate here on the latter.

The conversion from UV luminosity to SFR commonly used assumes a $>\,$100~Myr duration of star formation.  If this is not actually satisfied, SFR values may be higher by a factor of $\sim\!2$ \citep{Bouwens:2009at}.  Otherwise, one would expect $\dot{\rho}_*(z)$ to smoothly evolve over any $\gtrsim\,$100~Myr period.  This is a non-trivial amount of time when working in terms of $z$ (see the axes in Fig.~\ref{SFH}), especially at low mass where feedback may play a significant regulatory role.  The lifetimes of the massive stars that give rise to GRBs are much less than this duration; however, we find a GRB rate that remains consistent with the levels from $\sim\,$100~Myr earlier.  This is in contrast to the steep drop in the LBG-derived data in Fig.~\ref{SFH} between $z\!\approx\!8$ and $z\!\approx\!9$, a period of $\sim\,$100~Myr.  Even if $\dot{\rho}_*$ in the $z\!\sim\!9$ bin is increased by $\sim\! 2$, a discrepancy remains.  If the faint-end slope of the LF is steeper, say $\alpha\!\sim\!-2.2$, there would be enough faint galaxies for better agreement.  Beyond our last data point, we cannot yet constrain a possible drop in $\dot{\rho}_*$ at $z\!\gtrsim\!10$, though.  We discuss this regime further in Section~\ref{section:disc}.

These considerations also come into play when discussing the total integrated stellar mass density at high $z$.  \citet{Robertson:2011yu} showed that a simple integration of a cosmic SFR history roughly at the level of the GRB-inferred values exceeded the stellar mass densities in the range $4\!<\!z\!<\!8$ reported in \citet{Gonzalez:2010gx}.  However, the \citet{Gonzalez:2010gx} estimates rely on integrating their stellar mass functions only to the equivalent of a UV luminosity limit of $M_{\rm UV}\!=\!-18$.  It is clear that integrating to fainter limits would increase these values, reducing the tension with the GRB $\dot{\rho}_*$ (cf.\ \citealt{Stark:2012af}).  \citet{Robertson:2011yu} highlight other possibilities, including the influence of the cosmic background radiation at high $z$ and metallicity and IMF dependencies.  More directly, \citet{Wilkins:2013rt} showed that an evolving UV mass-to-light ratio results in steeper high-$z$ mass functions and more low-mass galaxies.

Another factor that may be involved is an evolving stellar IMF.  The relation between $\dot{\rho}_*$ and the stellar mass density is sensitive to the form of the IMF, while the GRB rate is not \citep{Wyithe:2009jc}.  A flatter (more top-heavy) IMF at higher $z$ has been invoked to explain the discrepancy between the star formation history and stellar mass density evolution at lower $z$ \citep{Wilkins:2008be}.  This raises the speculation that the population of faint, low-mass galaxies at $z \!>\! 7$ may be comprised disproportionately of very massive stars, and look very different from the population of low-$z$ galaxies.


\section{Discussion: To Reionization and Beyond}
\label{section:disc}

The extreme brightness of gamma-ray bursts and short lifetimes of their massive-stellar progenitors allow them, if calibrated, to serve as gauges of the instantaneous rate of star formation even in low-mass systems that are not detectable directly through starlight at present.  We have found that the $\dot{\rho}_*$ values resulting from GRB measurements, when examined in the context of recent high-$z$ LBG surveys, imply increasing numbers of faint galaxies at successively higher redshift.  This picture of the star formation history at high-$z$ agrees well with the sub-$L_*$, metal-poor GRB hosts observed at low $z$.  The rapid evolution of many faint dwarf systems to a smaller number of brighter systems suggests a combination of stellar mass growth through star formation combined with merging. This is, qualitatively at least, consistent with expectations from hierarchical growth.  Understanding this is crucial for studying the early evolution of galaxies (e.g., \citealt{Munoz:2010cn,Alvarez:2012nr,Ahn:2012sb,deSouza:2013wsa}).

We have found broad agreement with LBG observations by correcting for galaxies below detection thresholds, owing to the steeper faint-end slopes from recent surveys, agreeing with the general conclusion of \citet{Kistler:2009mv}, \citet{Bouwens:2011xu}, and \citet{Robertson:2013bq} that the level of star formation was sufficient to reionize the universe.  The specific implications for reionization are similar to those discussed in \citet{Kistler:2009mv} and \citet{Wyithe:2009jc}, which found that the ionizing photon flux was likely sufficient to achieve reionization by $z\!\sim\!8$.  In particular, the slow evolution is consistent with the optical depth to Thomson scattering of the CMB ($\tau\!=\!0.089\pm0.014$) and reionization redshift ($z_{\rm reion}\!=\!10.6\pm1.1$) inferred from {\it WMAP} polarization data \citep{Hinshaw:2012fq} with reasonable escape fractions.  {\it Planck} temperature/lensing data alone yield $\tau\!=\!0.089\pm0.032$ \citep{Planck} and polarization results are awaited.

This leaves the question of what occurred before the earliest recorded GRB.  \citet{Coe:2012wj}, based on the observation of a lensed, $M\!\approx\!-19.5$ galaxy at $z\!\sim\!11$, estimated the required UV LF in this range.  Integrating this down to $M\!\sim\!-19$ gives the lower result in Fig.~\ref{SFH}.  If we instead use $M_{\rm cut}\!\sim\!-10$, as at $z\!\sim\!8$, we arrive at the upper $z\!\sim\!11$ $\dot{\rho}_*$ value in Fig.~\ref{SFH}.  We see that this is not too dissimilar from $z\!\sim\!9$, and may also need to be scaled up.  This is consistent with a slowly-declining $\dot{\rho}_*(z)$ and hints that $z\!\gtrsim\!10$ GRBs may be as prevalent as at $z\!\sim\!8\!-\!9$.  The observation of even one would carry major significance.

Our results suggest a larger amount of star formation in faint galaxies than could ultimately be maintained, although for quantitative comparisons we have used the latest UV luminosity function fits.  However, it is not clear how well extrapolations of a Schechter function actually represent the faint galaxy contribution.  Moreover, the highest-$z$ fits rely on measurements at lower $z$ and/or empirical relations.  In this rapidly-evolving era, with changes occurring over a variety of scales with various origins, the accuracy of a scale-independent function can be questioned.  Recent studies have suggested that decreased H$_2$ formation in low-metallicity galaxies in this epoch could suppress star formation rates at low masses \citep{Kuhlen:2011dt,Jaacks:2013hx}.

In particular, the simulations of \citet{Jaacks:2013hx} display a $z$-dependent break in the UV LF function at the faint end.  This provides a physically-motivated limit to UV LF extrapolations and highlights the limitations of using a fixed magnitude cutoff in converting measurements of the UV LF to a total $\dot{\rho}_*$ at high $z$.  As simulations improve, we can acquire a better understanding of the relation between the GRB- and UV-based $\dot{\rho}_*$ values and whether there exists a need for new channels for forming stars and/or gamma-ray bursts.

Since each GRB arises from a single star in a single galaxy, our discussion thus far at the highest redshifts has necessarily been limited to determining the $\dot{\rho}_*$ needed to make the observed event rates likely.  However, this discretization provides a distinct advantage, as the combination of a steep UV LF and a sharp break would result in a peak in the SFR distribution, with $\dot{\rho}_*$ dominated by galaxies of a similar SFR.  A high-$z$ GRB would most likely arise from such a galaxy, regardless of the exact calibration from GRB rate to $\dot{\rho}_*$.  Since a galaxy is guaranteed to be at the position of each GRB, deep searches could thus use these as short cuts to determine this characteristic scale at the measured redshift of a given GRB.

\acknowledgments
We thank Nat Butler for his updated GRB data.  We acknowledge use of the \textit{Swift} public archive.
MDK acknowledges support provided by NASA through the Einstein Fellowship Program, grant PF0-110074 and
HY during a visit to Berkeley by US DOE contract DE-SC00046548.

\clearpage

\end{document}